\documentclass[preprint,prb,showpacs,color,superscriptaddress,epsfig]{revtex4}
\usepackage{graphicx}
\usepackage{epsfig}
\usepackage{amsmath}
\usepackage{epstopdf}
\usepackage{array,subfig}
\begin{document}
\title{Electronic structure and Jahn-Teller effect in GaN:Mn and ZnS:Cr}
\author{F. Virot}
\affiliation{Institut Mat\'eriaux Micro\'electronique Nanosciences de Provence,
Facult\'e St. J\'er\^ome, Case 142, F-13397 Marseille Cedex 20, France}
\author{R. Hayn}
\affiliation{Institut Mat\'eriaux Micro\'electronique Nanosciences de Provence,
Facult\'e St. J\'er\^ome, Case 142, F-13397 Marseille Cedex 20, France}
\author{A. Boukortt}
\affiliation{Institut Mat\'eriaux Micro\'electronique Nanosciences de Provence,
Facult\'e St. J\'er\^ome, Case 142, F-13397 Marseille Cedex 20, France}
\affiliation{ Mostaganem University, Faculty of Science and Technology, Genie
Electric Department 27000, Algeria}
\date{\today}

\begin{abstract}
We present an ab-initio and analytical study of the Jahn-Teller
effect in two diluted magnetic semiconductors (DMS) with
$d^4$ impurities, namely Mn-doped GaN and Cr-doped ZnS.
We show that only the combined treatment of Jahn-Teller distortion and
strong electron correlation in the $3d$ shell may lead to the correct
insulating electronic structure. Using the LSDA+$U$ approach
we obtain the Jahn-Teller energy gain in reasonable agreement
with the available experimental data. The ab-initio results
are completed by a more phenomenological ligand field theory.
\end{abstract}


\maketitle

\section{Introduction}
The Jahn-Teller effect of impurity ions in semiconductors with
degenerate ground state is well known since long time. Most of its
theoretical treatments were based on crystal field theory. Actually,
the class of diluted magnetic semiconductors regains a lot of
interest in connection with the search for new materials for
spintronics applications. For example, high Curie temperatures were
predicted in GaN:Mn. \cite{Dietl01} Corresponding experimental
\cite{Hidenobu02} or ab-initio studies \cite{Sato03} seemed to confirm these
predictions. However, the experimental results are very disputed
since they might be caused by small inclusions of secondary phases.
\cite{Sarigiannidou06} And also most of the previous ab-initio
studies obtained a partially filled band of only one spin direction
at the Fermi level (half-metallic behavior) which will be shown to
be an artifact of those calculations. Actually, the Mn ion changes
its valence in the chemical series from GaAs:Mn, via GaP:Mn to
GaN:Mn. \cite{Schulthess06} It is Mn$^{2+}$ in GaAs:Mn (for a
sufficiently high Mn-concentration)\cite{Jungwirth07} which leads to
hole doping, but it remains Mn$^{3+}$ in GaN:Mn. And the Jahn-Teller
effect is crucial to stabilize Mn$^{3+}$. As will be shown below, only the combined
treatment of Jahn-Teller effect and strong electron correlations
leads to the correct electronic structure. The electron correlations
turn out to be the leading interaction, but the Jahn-Teller effect
is necessary to break the symmetry.
%
%
%
%in GaN:Mn were reported experimentally and confirmed by numerous
%ab-initio calculations. However, those experimental results are very
%disputed and might be caused by secondary phases.

We present here a combined ab-initio and crystal field theory of
magnetic ions in II-VI or III-V semiconductors. As representative
examples we treat the 3$d^4$ ions Mn$^{3+}$ in GaN and  Cr$^{2+}$ in
ZnS. The first one is chosen because of its actual interest for
spintronic applications and the second one since it is a very well
studied model system for the Jahn-Teller effect of $d^4$ ions.
\cite{Vallin70} Our ab-initio results are in good agreement
with the experimental data, but only if we include properly the
effects of the electron correlations in the 3$d$ shell. For that
purpose we use the LSDA+$U$ method. The resulting electronic
structure corrects previous electronic structure calculations (which
did not take into account the combined effect of Jahn-Teller
distortion and Coulomb correlation, neither for Mn-doped GaN
\cite{Sanyal03,Kulatov02} nor for Cr-doped ZnS
\cite{McNorton08,Tablero06}) in a dramatic way: instead of
half-metallic  behavior we obtain an insulating ground state with a
considerable excitation gap. Similar results for GaN:Mn were
reported earlier but using different methods than in our study.
\cite{Schulthess06,Stroppa09} In a second step, to make contact with
the traditional literature on that subject, we connect our ab-initio
results with crystal field theory. We obtain the complete set of
crystal field parameters in good agreement with previous optical
measurements.\cite{Vallin70,Kaminska79,Wolos04}

We treat the host crystals in zinc-blende phase. Both magnetic ions
are in the 3$d^4$ configuration. The  electronic level is split by a
cubic crystal field created by the first nearest neighbors (ligands)
of the transition metal ion. But it remains partially filled and the
Jahn-Teller effect may occur which induces the splitting of the
partially filled level due to displacements of the ligands around
the transition metal ion. The local  symmetry of the crystal is
reduced and the total energy of the supercell is minimized. The
energy gain is denoted as $E_{JT}$.

\section{Method of calculation}
Our calculations are performed using the full potential local
orbital (FPLO)\cite{Koepernik99}  method with the LSDA+$U$ (local spin density
approximation with strong Coulomb interaction)\cite{Anisimov91} approximation in the
atomic limit scheme. The lattice constants are optimized for the
pure semiconductors using the LSDA method, $a_0=4.48$ \AA \ for GaN
and $a_0=5.32$ \AA \  for  ZnS.  To study the Jahn-Teller effect, we
use a supercell of 64 atoms in zinc blende phase, and a
$4\times4\times4$ k-point mesh. The LSDA+$U$ parameters are introduced
as Slater parameters : for  Mn (see Ref.\ \onlinecite{Anisimov91}),
$F^2=7.41$ eV, $F^4=4.63$ eV and for  Cr (see
Ref.\ \onlinecite{Korotin98}), $F^2=7.49$ eV, $F^4=4.68$ eV. The $F^0$
parameter is equal to $U$, and is chosen to be 4 eV. This value is
in good agreement with  other works and it corresponds to the value
which gives the maximum  splitting of the triplet state in $D_{2d}$
symmetry. We have verified that our results are not very sensitive
to the actual choice of the $U$ parameter. The value $U=4$ eV gives
representative results for a rather large range of $U$ parameters
reaching from 3 up to 8 eV.
\begin{figure}[h]
%\begin{center}
\includegraphics[scale=0.12]{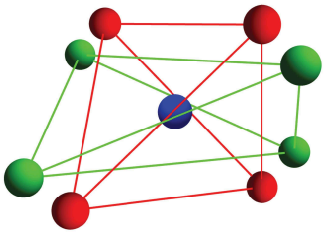}
\caption{(Color online) Schematic drawing of a tetragonal distortion  
from $T_d$ (red spheres) to $D_{2d}$ (green spheres).}\label{f1}
%\end{center}
\end{figure}

The transition metal ions (Mn or Cr) substitute one atom in the
center of the crystal and there is a complete
tetrahedron of ligands around the magnetic ion. Without distortion,
the tetrahedron is in the $T_{d}$ symmetry group (cubic). By symmetry,
there are two tetragonal (point groups $C_{2v}$ and $D_{2d}$) and
one trigonal ($C_{3v}$) possible Jahn-Teller distortions. We have
verified that the most important energy gain is obtained with the
pure tetragonal distortion where the symmetry reduces from $T_{d}$
to $D_{2d}$. That is in agreement with a previous study of the
Jahn-Teller effect in GaN:Mn where, however, the electron
correlation had not been taken into account. \cite{Luo05} The
schematic displacement of ligands is represented in  Fig. \ref {f1},
it is defined by $\delta_x=\delta_y \neq \delta_z$.

\section{Ab-initio results}
\begin{figure}[h]
%\begin{center}
\subfloat[GaN:Mn]{%
\includegraphics[scale=0.30]{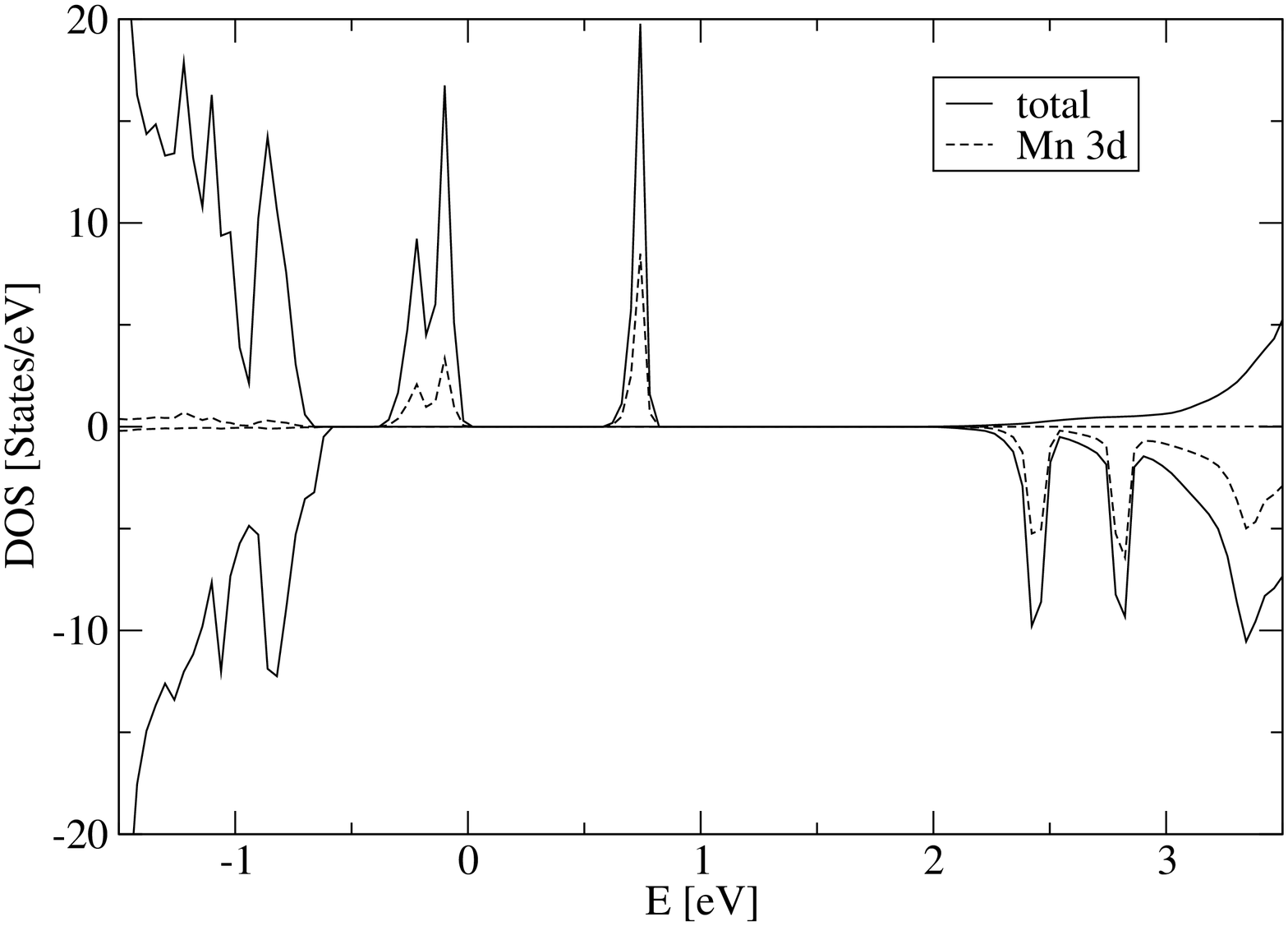}}
\hspace{1cm}
\subfloat[ZnS:Cr]{%
\includegraphics[scale=0.30]{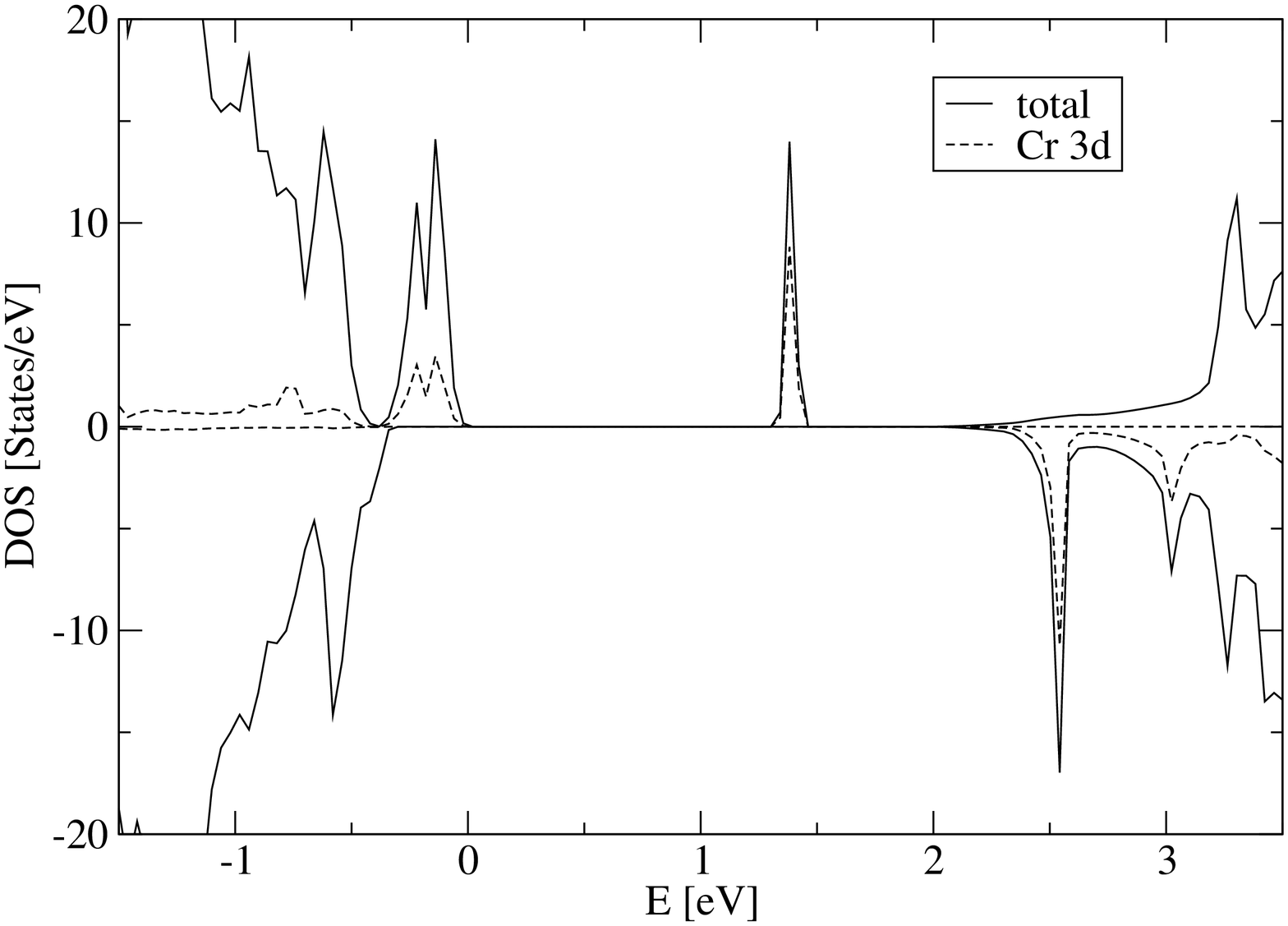} }
\caption{Partial density of states (DOS) of the two DMS. The results
were obtained with the LSDA+$U$ method ($U=4$ eV). The dashed line represents
the partial $3d$ DOS 
of the transition metal ion and the solid line represents the total DOS. The Fermi level 
is set to 0 eV.}\label{f2}
%\end{center}
\end{figure}

To study the Jahn-Teller distortion we use a supercell of 64 atoms. We 
performed a series of calculations with different displacements of the
ligands around the impurity in order to find the configuration which
minimizes the total energy of the supercell.  The preferred configuration 
is of $D_{2d}$ symmetry and the results  for the
density of states (DOS) are presented in Fig. \ref{f2}. For
Mn-doped GaN, the distance between Mn and N is  $1.942$ \AA  \ (with
cubic symmetry, it is $1.937$ \AA) with $\delta_x=1.68$ pm and
$\delta_z=1.76$ pm. The Jahn-Teller effect induces a lowering of
the  total energy by  38.13 meV and a splitting of the triplet state
by 0.81 eV. For Cr-doped ZnS, the results are similar, the distance
Cr-S is $2.314$ \AA \ ($2.304$  \AA \ without Jahn-Teller effect),
with $\delta_x=3.36$ pm and $\delta_z=-0.32$ pm. The total energy
decreases by $58.1$ meV and the splitting of the triplet level is
$1.44$ eV. 

Without $U$, in the LSDA method, we also find a Jahn-Teller effect for 
GaN:Mn but the energy gain is smaller by nearly two orders of magnitude,
only 0.71 meV. In the LSDA method all the 3$d$ levels are located in 
the gap with a rather small mixing to the $2p$ orbitals. Without Jahn-Teller 
effect, there is a small band of $t_{2g}$ character which is partially filled
and one finds half-metallic behavior. 
The Hubbard correlation opens up a considerable gap. 
The empty state (singlet) is mainly of 3$d$
character for the two compounds and the doublet, just below the Fermi
level is mainly of 2$p$ character originating from ligand orbitals.
That change of orbital character arises since the LSDA+$U$ method
pushes the occupied 3$d$ levels much lower in energy than in the 
LSDA method. As a consequence, the transition to the first 
excited state has a considerable $p$-$d$ character and should be visible as
an optical interband transition. That allows a reinterpretation of the
optical transition at 1.4 eV (for GaN:Mn) which is usually considered as a pure
$d$-$d$ transition. \cite{Wolos04} In agreement with a proposal of Dietl it 
corresponds to a transition from the $d^4$ configuration to $d^5$ and ligand 
hole. \cite{Dietl08}

There are two peaks in the unoccupied, minority DOS at about 2.5 eV for GaN:Mn. They correspond to crystal field split 3$d$ levels which were well seen in X-ray absorption spectroscopy. Up to now, they were interpreted by means of a LSDA calculation, \cite{Titov05} but our DOS shows that these peaks occur also in the more realistic LSDA+$U$ approach where, however, a detailed calculation of the matrix element effects (which was performed in Ref.\ \onlinecite{Titov05}) is still lacking.

%These results show that the  compound is now in an
%insulating state. The empty state (singlet) is mainly of 3$d$
%character in the two cases and the doublet, just below the Fermi
%level is mainly of 2$p$ character originating from ligand orbitals.
%A possible interpretation  is that the transition to the first
%excited state is of $p$-$d$ type, whereas it was interpreted to be a
%$d$-$d$ transition in most previous publications.\cite{Titov05}

For GaN:Mn, the total magnetic moment is equal to $4 \mu_B$, corresponding
 to $S=2$. That fits well with the ionicity $3+$ for manganese. The local 
 magnetic moment at the manganese site is slightly enhanced 
 $M_{Mn}=4.042 \mu_B$ which is compensated by small induced 
 magnetic moments of opposite sign at the neighboring ligands
 and further neighbors. 

\begin{figure}[h]
%\begin{center}
\subfloat[GaN:Mn]{%
\includegraphics[scale=0.30]{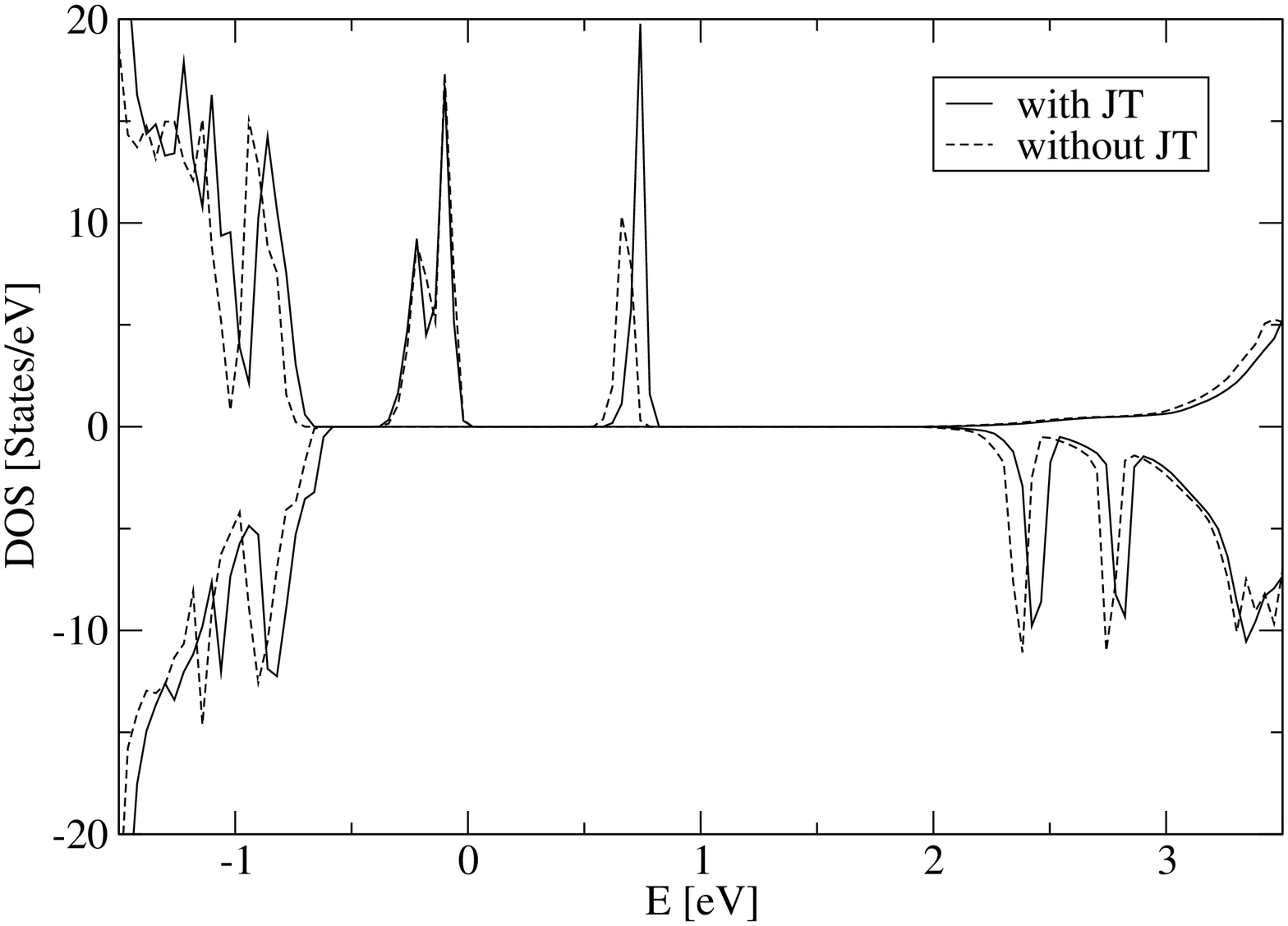}}
\hspace{1cm}
\subfloat[ZnS:Cr]{%
\includegraphics[scale=0.30]{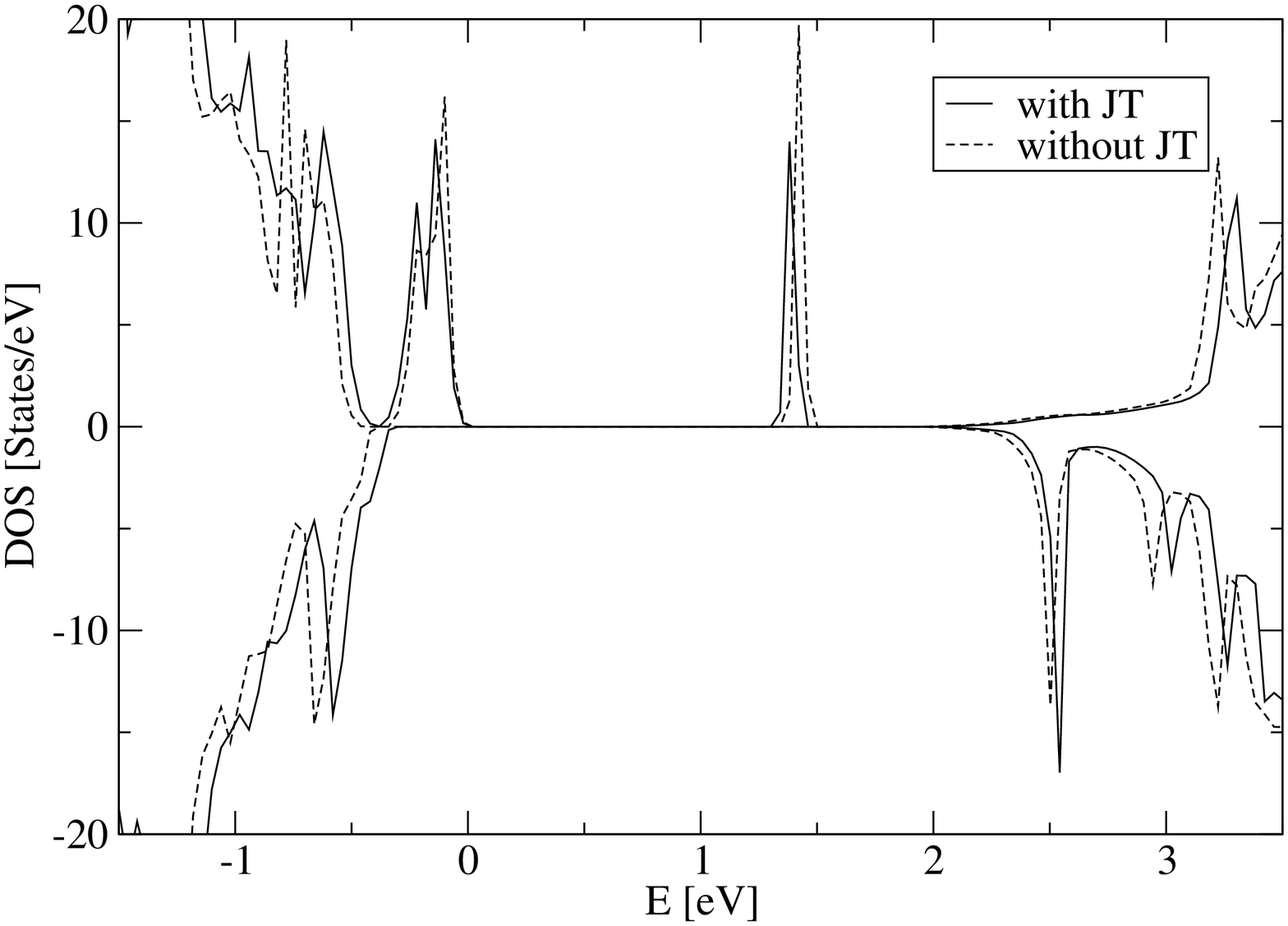} }
\caption{DOS resulting from the LSDA+$U$ calculation ($U=4$ eV). The dashed
line represents the DOS without Jahn-Teller distortion, and the solid
line represents the DOS with Jahn-Teller effect but by breaking the 
local cubic symmetry. The Fermi level is set
to 0 eV. }\label{f3}
%\end{center}
\end{figure}

Another  interesting  result is presented  in  Fig. \ref{f3}. In
this case we introduced the local Coulomb correlation but no lattice
deformation. The local cubic symmetry was however broken by the
occupation of the triplet state. The calculation still shows an
insulating case with nearly the same splitting of the triplet state.
This observation is not visible within the LSDA method. Therefore,
we can say that the splitting of the triplet state is mainly due to
the strong correlation of 3$d$ electrons. The Jahn-Teller distortion
introduces a small additional splitting of the triplet state : $60$
meV for Mn-doped GaN and $-11.4$ meV for Cr-doped ZnS. This interpretation is
confirmed by a pure LSDA calculation, because it gives an identical
value of the splitting with the same ligand coordinates. A negative
value of the splitting corresponds  to a stretching of the tetrahedron
along the z axis (Fig. \ref{f3}(b)).

\section{Ligand field theory}
\begin{figure}[h]
%\begin{center}
\includegraphics[scale=0.17]{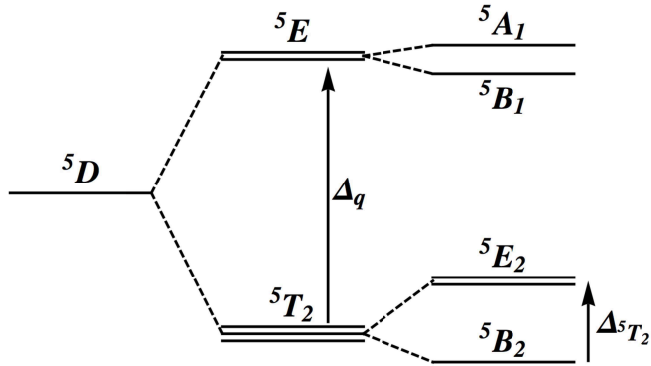}
\caption{Schematic multiplet spectrum for a transition metal ion in
$d^4$ configuration. From left to right, $^5D$: isolated ion,
cubic crystal field splitting, and level splitting due to the Jahn-Teller distortion.
}\label{f4}
%\end{center}
\end{figure}

For a deeper understanding of the Jahn-Teller effect we treat it also in 
ligand field theory. In that theory, the degeneracy of the impurity 3$d$ level 
is split due to hybridization with the neighboring ligands. In the following we 
neglect the electrostatic contributions which will be shown to be 
justified for GaN:Mn and to a lesser extent for ZnS:Cr. When the 
3$d$ ion is in the center of an ideal tetrahedron,
the cubic  crystal field  splits this level into a triplet state and
a doublet state. The distance separating these two levels is denoted
as $\Delta_{q}$. Then, the Jahn-Teller effect splits the doublet state
into two singlet states, and the triplet state into a doublet state
and a singlet state. The distance between these levels is denoted
as $\Delta_{^5T_2}$ (shown in Fig. \ref{f4}). The local Hamiltonian in 
ligand field or crystal field theory can be expressed as
\begin{equation}
\begin{split}
H_{CF} & =H_{cub}+H_{tetra} \\
& =B_{4}(O^{0}_{4}+5O^{4}_{4})+(B^{0}_{2}O^{0}_{2}+B^{0}_{4}O^{0}_{4})
\end{split}
\label{eq1}
\end{equation}
where $B_k^q$ and $O_k^q$ are Steven's parameters and  Steven's
operators, respectively. \cite{Abragam70} The first part represents the 
Hamiltonian of an ideal tetrahedron and the second describes the linear Jahn-Teller
effect. The eigenvalues of $H_{CF}$ correspond to the 3$d$
electronic levels of the magnetic ion (they represent the spectrum).
$B_4$ is the parameter of the cubic crystal field, and it is equal to
$\Delta_{q}/120$.

\begin{figure}[h]
%\begin{center}
\includegraphics[scale=0.1]{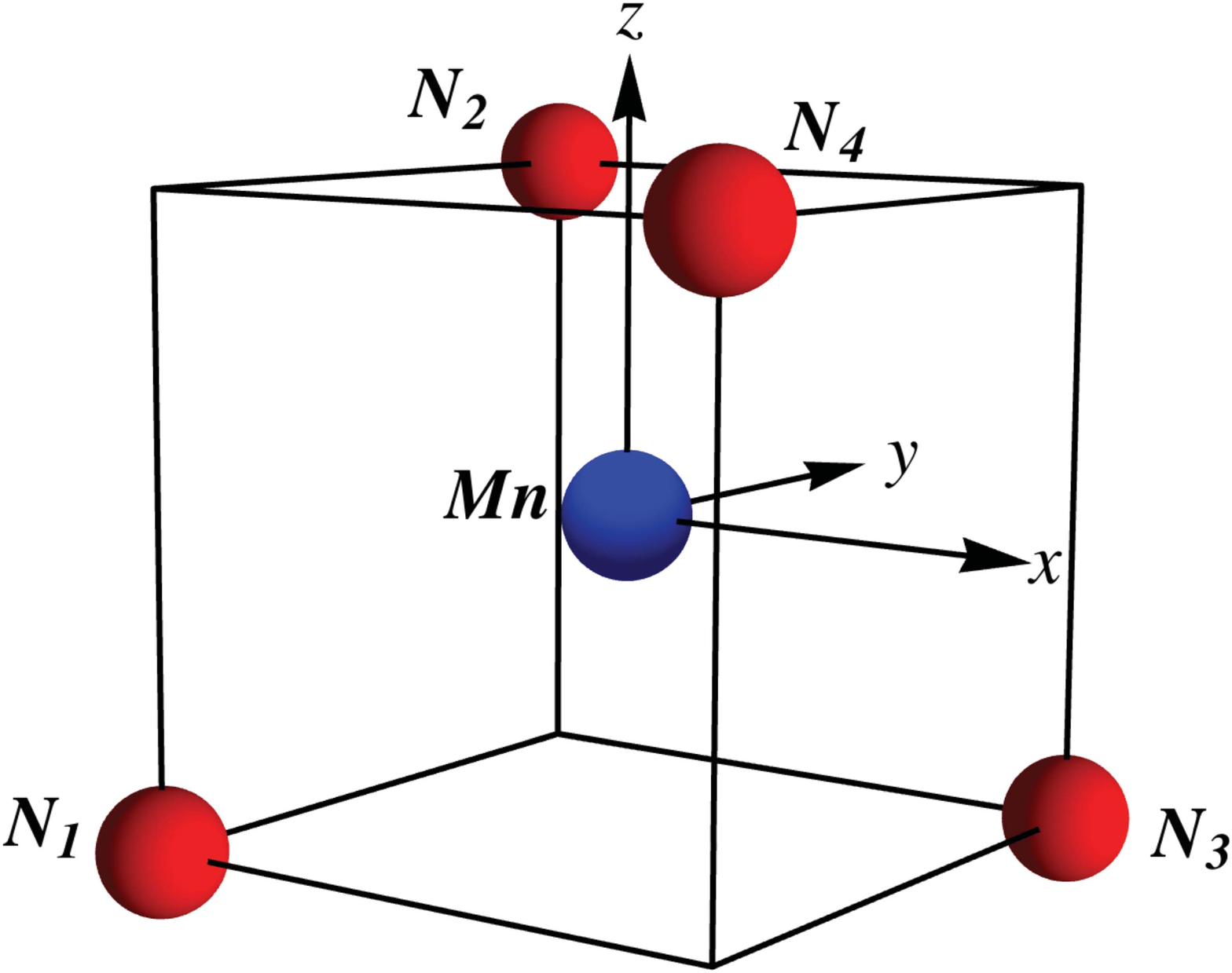}
\includegraphics[scale=0.1]{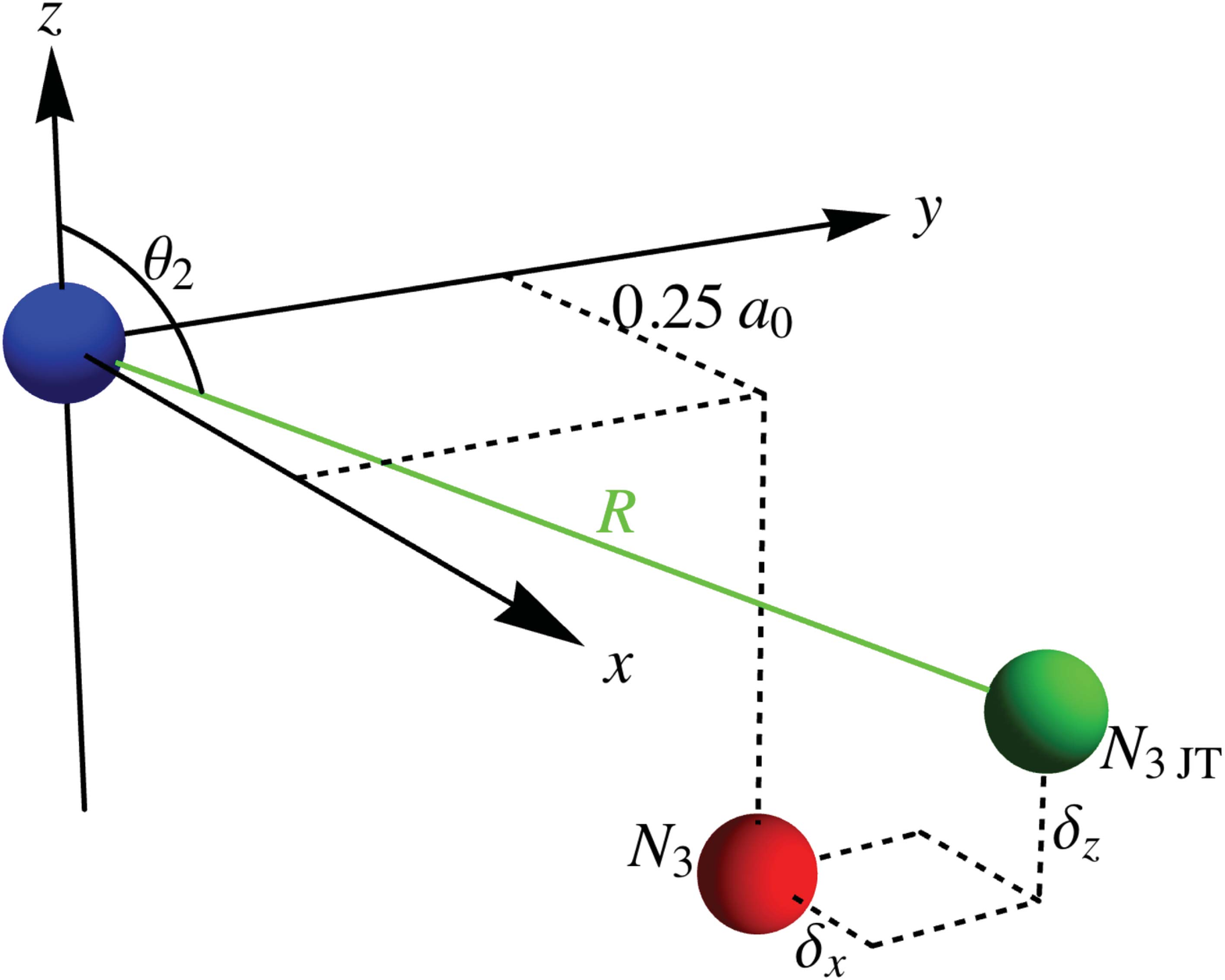}
\caption{(Color online) Definition of coordinate axes and lattice displacements used 
in the present work. Example of MnN$_4$. }\label{f5}
%\end{center}
\end{figure}

The crystal field Hamiltonian (\ref{eq1}) has the same form for the 
one-particle problem (with parameters $B_4$, $B_2^0$, and $B_4^0$) 
or for the 3$d^4$ multiplet (with parameters 
$\tilde{B}_4$, $\tilde{B}_2^0$, and $\tilde{B}_4^0$). \cite{Abragam70,Kuzian06}
The fundamental
multiplet state is $^5D$ ($L=2$ and $S=2$). In this fundamental
state, the orbital moment for one electron ($l=2$) is equal to the total
orbital moment. This particular case induces that the one electron
spectrum is opposite to the multiplet spectrum (ex :
$\tilde{B}_4=-B_4$). In the superposition model \cite{Kuzian06,Bradbury67,Kuzmin91} the
crystal field Hamiltonian (\ref{eq1}) can be calculated by adding up the hybridization
contributions of all the ligands. Then, the matrix elements of $H_{CF}$ with respect to the 
projection $m_L$ of the total orbital momentum $L=2$ are given by
\begin{equation}
\begin{split}
V_{m_{L},m'_{L}} =& \sum_{i}\left[ A_{m'_{L},m_{L}}b_{4}(R_{i})Y_{4}^{m_{L}-m'_{L}}(\theta_{i},\phi_{i})\right]\\
 &+\sum_{i}\left[B_{m'_{L},m_{L}}b_{2}(R_{i})Y_{2}^{m_{L}-m'_{L}}(\theta_{i},\phi_{i})\right]\\
 &+\sum_{i}\left[b_{0}(R_{i})\delta_{m'_{L},m_{L}} \right] \\
\end{split}
\end{equation}
with 
\begin{equation}
\begin{split}
& A_{m'_{L},m_{L}}=\frac{(-1)^{m'_L}5\sqrt{4\pi}}{27}C^{224}_{-m'_{L},m_{L}}C^{224}_{0} \\
& B_{m'_{L},m_{L}}=\frac{(-1)^{m'_L} \sqrt{4\pi}}{\sqrt{5}}C^{222}_{-m'_{L},m_{L}}C^{222}_{0}\\
\end{split}
\end{equation}
and where $C_{m_{1}m_{2}}^{j_{1}j_{2}J}$ are Clebsch-Gordon
coefficients and $Y_{L}^{m_L,m_{L'}}$ are  spherical harmonics. The
$i$ index corresponds to the ligand number. The axis system  is defined in Fig. \ref{f5}.

In identifying the eigenvalues of $H_{CF}$ with the equivalent values
of $V_{m_{L},m'_{L}} $, we obtain the expressions of the Steven's
parameters as a function of the ligand coordinates
\begin{equation}
\begin{split}
&\tilde{B}_4=\frac{-b_4(R)\tilde{Y}_{4}^{4}(\theta_2,\theta_3)}{120}\\
&\tilde{B}_{4}^{0}=\frac{b_4(R)}{84}\left(\frac{\tilde{Y}_{4}^{0}(\theta_2,\theta_3)}{12}
+\frac{7\tilde{Y}_{4}^{4}(\theta_2,\theta_3)}{12}\right)\\
&\tilde{B}_2^0=\frac{-2b_2(R)\tilde{Y}_{2}^{0}(\theta_2,\theta_3)}{49}
\end{split}
\end{equation}
with 
\begin{equation}
\begin{split}
&\tilde{Y}_{4}^{0}(\theta_2,\theta_3)=6-30(\cos^2\theta_2+\cos^2\theta_3)+35(\cos^4\theta_2+\cos^4\theta_3)\\
&\tilde{Y}_{4}^{4}(\theta_2,\theta_3)=\sin^4\theta_2+\sin^4\theta_3\\
&\tilde{Y}_{2}^{0}(\theta_2,\theta_3)=-2+3\cos^2\theta_2+3\cos^2\theta_3\\
&b_2(R)=\frac{\hbar^4R^3_d}{ \Delta_{pd}m^2R^7}(\eta^2_{pd\sigma}+\eta^2_{pd\pi})\\
&b_4(R)=\frac{9\hbar^4R^3_d}{5 \Delta_{pd}m^2R^7}(\eta^2_{pd\sigma}-\frac{4}{3}\eta^2_{pd\pi})\\
\end{split}
\end{equation}
where  $b_2(R)$ and $b_4(R)$ are expressed by means of the Harrison
parametrization of $p$-$d$ hopping. The values of $\eta_{pd\sigma}=-2.95$ and 
$\eta_{pd\pi}=1.36$
are extracted from his book (first edition). \cite{Harrison80} 
In the case of $D_{2d}$ symmetry, the distances  $R$, between
magnetic ion and the four ligands are identical. $\Delta_{pd}$ is
the charge transfer energy. It is treated as an adjustable parameter
in this theory.

\begin{table}[h]
\caption{Summary of parameters.}\label{t1}
\begin{tabular}{ccccc}
\hline \hline
&\multicolumn{2}{c}{GaN:Mn}&\multicolumn{2}{c}{ZnS:Cr}\\
&our work&experiment\footnotemark[1]&our work&experiment \\ \hline
$\Delta_{^5T_2}$ (meV)&187&111&167.1&213.9\footnotemark[2]  111.6\footnotemark[3] \\
$E_{JT}$ (meV)&59.9&37&58.4&71.3\footnotemark[2] , 37.2\footnotemark[3] \\
$\tilde{B}_4^0$ (meV)&-1.88&-1.05&-1.26&\\
$\tilde{B}_2^0$ (meV)&-8.24&-3.98&-10.15&\\
$\Delta_q $(eV)&1.4&1.37&0.57&0.58\footnotemark[2] , 0.59\footnotemark[3] \\
$R$ (\AA)&1.932&&2.303&\\
$\delta_x$ (pm)&1.28&&1.65&\\
$\delta_z$ (pm)&3.79&&3.46&\\ \hline
$\Delta_{pd}$ (eV)&2.31&&& \\
$b_2/b_4$&&&3.3&\\ \hline \hline
\end{tabular}
\footnotetext[1]{optical masurements for wurtzite GaN:Mn, Wolos {\em et al}, 
Ref.\  ~\onlinecite{Wolos04}.}
\footnotetext[2]{optical measurements, Vallin {\em et al}, Ref.\  ~\onlinecite{Vallin74}.}
\footnotetext[3]{optical measurements, Kaminska {\em et al}, Ref.\ ~\onlinecite{Kaminska79}.}
\end{table}

Hamiltonian (\ref{eq1}) concerns the electronic degrees of freedom only. The 
tetragonal distortion results from the coupling to the lattice. As it is shown in Fig.\
\ref{f4} any tetragonal distortion leads to a splitting of the lowest triplet $^5T_2$ and
to an energy gain. This energy gain is linear in the lattice displacements, whereas the 
vibronic energy loss  $E_{vibronic}$ is a quadratic term. One has to minimize 
the total energy
\begin{equation}
\Delta_{total}=\frac{2}{3}\left(E(^5B_2)-E(^5E_2)\right)+E_{vibronic} \;  .
\end{equation}
For the sake of simplicity we approximate
$E_{vibronic}$ by the breathing mode energy, extracted from the LSDA+$U$ calculation.
We find the lattice contribution for GaN:Mn to be $E_{vibronic} [\mbox{eV}]=36.7708 R'^2$ and for ZnS:Cr 
$E_{vibronic} [\mbox{eV}]=30.4887 R'^2$  with
$R'=\sqrt{\delta_x^2+\delta_y^2+\delta_z^2}$ in \AA. One should note, that 
there is no difference between LSDA and LSDA+$U$ 
methods for 
the lattice energy. Also, we have probed the lattice energy for the more specific tetragonal 
mode with no essential difference. The energy gain which is induced by the Jahn-Teller 
effect, corresponds to :
\begin{equation}
E_{JT}= min(\Delta_{total}) \; .
\end{equation}

\begin{table*}
\caption{Comparison of parameters. When possible, the complete set of parameters was calculated from the literatures values by the relations given in Sect V. HSE : Heyd-Scuseria-Ernzerhof hybrid functional, GGA : Generalized Gradient Aproximation, GFC : Green-function Calculation.} \label{t2}
\begin{ruledtabular}
\begin{tabular}{ccccccc}
&&method&$E_{JT}$ [meV]&$Q_{\theta}$ [\AA]&V [eV/\AA]&$\hbar \omega$ [$cm^{-1}$] \\ \hline
GaN:Mn&present work&ligand field theory&59.9&-0.0828&-1.44&579 \\
&&ab-initio, LSDA+$U$&38.13&-0.0562&-1.35&680.5 \\
&literature values&ab-initio, GGA\footnotemark[1]&100&-0.1365&-1.46&454\\ 
&&ab-initio, HSE\footnotemark[2]&184&&&\\
&&experiment, optics\footnotemark[3]&37&&&\\
\hline 
ZnS:Cr&present work&ligand field theory&58.4&-0.0834&-1.4&375.8\\
&&ab-initio, LSDA+$U$&58.4&-0.0496&-2.35&631.5\\
&literature values&ab-initio, GFC\footnotemark[4]&185.98&-0.16&-2.32&349.4\\ 
&&experiment, optics\footnotemark[5]&37.2&-0.279&-0.266&90\\
\end{tabular}
\end{ruledtabular}
\footnotetext[1]{Luo {\em et al.} Ref.  ~\onlinecite{Luo05}.}
\footnotetext[2]{Stroppa {\em et al.} Ref.  ~\onlinecite{Stroppa09}.}
\footnotetext[3]{Wolos {\em et al.} Ref.  ~\onlinecite{Wolos04}, wurzite GaN:Mn.}
\footnotetext[4]{Oshiyama {\em et al.} Ref.  ~\onlinecite{Oshiyama88}.}
\footnotetext[5]{Kaminska {\em et al.} Ref.  ~\onlinecite{Kaminska79}.}
\end{table*}

The results of the ligand field model are presented in Table \ref{t1}. 
For GaN:Mn, we have exclusively used the ligand hybridization as the microscopic origin
for the level splitting. The value of $\Delta_{pd}$ is adjusted such  
that $\Delta_q$ equals the experimental value. The results are very convincing 
which proves that the exceptional large value of $\Delta_q=1.4$  eV in the case of 
GaN:Mn is dominantly caused by the hybridization energy to the ligands. 
On the other hand, the neglect of the electrostatic 
contribution is certainly an approximation which  shows its limits for ZnS:Cr. The 
procedure described above gives no satisfactory results. We interpret this deficiency
such that the electrostatic corrections become more important for ZnS:Cr which has a 
much smaller value of $\Delta_q$ indicating a smaller hybridization. As discussed 
in Ref.\   \onlinecite{Savoyant09}, the higher order crystal field parameters, and 
especially  $\tilde{B_4^0}$ are certainly more influenced by further reaching neighbors
than $\tilde{B_4}$. Therefore,  we determine the $b_4$ parameter from the 
experimentally known value of $\Delta_q$ and 
introduce a second free parameter ($b_2/b_4$) which
is fitted to the LSDA+$U$ energy gain. The parameter set for ZnS:Cr which 
was found in such a manner (Table \ref{t1}) is now in good agreement with previous
optical measurements. 

\section{Comparison to other works}
The lattice displacements obtained from the ligand field theory (Table \ref{t1}) fulfill 
approximately the relationship $\delta_x=\delta_y=\frac{1}{2}\delta_z$. That occurs 
not by accident 
since this relationship characterizes a pure tetragonal mode $Q_{\theta}$. In general, 
up to now, we 
considered $\delta_x=\delta_y\neq \delta_z$, which corresponds to a mixture of tetragonal 
and breathing mode $Q_b$ ($\delta_x=\delta_y=- \delta_z$). 
The normal coordinates of the two modes are defined as  
$Q_{\theta}= - 2\sqrt{\frac{2}{3}}(\delta_x+\delta_z)$ and 
$Q_b=\frac{2}{\sqrt{3}}(2\delta_x-\delta_z)$, respectively. \cite{Sturge67} 
But it is only the tetragonal mode
which leads to a splitting of the  $^5T_2$ level. That is described in the Hamiltonian which 
was used by Vallin {\em et al} $\;$  \cite{Vallin70, Vallin74,Kaminska79} to analyze their 
data obtained by optical 
measurements and 
electron paramagnetic resonance (EPR)  
\begin{equation}
 H_D=VQ_\theta \epsilon_{\theta}+\frac{\kappa}{2}Q^2_{\theta} \; . 
\label{eq8}
\end{equation}
Here, the $V$ parameter is the Jahn-Teller coupling coefficient, and 
$\epsilon_{\theta}$ is a diagonal 3*3 matrix which describes the splitting of the $^5T_2$ 
triplet into the upper $^5E_2$ doublet and the lower $^5B_2$ singlet. Its diagonal elements 
are  1/2 (corresponding to $^5E_2$) and -1 (corresponding to$^5B_2$). The 
parameter $\kappa$
describes the lattice stiffness and is connected with the phonon frequency by 
$\omega =  \sqrt{\kappa/m}$ where $m$ is the mass of one ligand. Minimizing the
energy (\ref{eq8}) we find the relationship $V=\kappa Q_{\theta}$ and the Jahn-Teller energy
%With $Q_{\theta}=2\sqrt{\frac{2}{3}}(\delta_x+\delta_z)$, the normal coordinate 
%described from Sturge's work.\cite{Sturge67} These equations presented here are 
%valid only for a pure tetragonal distortion ($\delta_x=\delta_y=\frac{1}{2}\delta_z$). In 
%his case, the Jahn-Teller energy is equal :
\begin{equation}
E_{JT}=\frac{V}{2}Q_{\theta} \; .
\label{for1}
\end{equation}
Therefore, if we know the Jahn-Teller energy $E_{JT}$ and the lattice distortion $Q_{\theta}$, 
we may calculate the coupling coefficient $V$ and the phonon frequency 
$\omega$ and compare it with other works. The comparison is not an ideal one since our
LSDA+$U$ results do not correspond to a pure tetragonal distortion. They contain an important
part of the breathing mode which comes about due to recharging effects around the impurity
being not treated in the ligand field theory or in the analysis of the optical data in 
Refs.\  \onlinecite{Vallin70} and \onlinecite{Kaminska79}. Nevertheless,  we compare 
in Table \ref{t2} our ab-initio and ligand field results
with other theoretical or experimental work from the literature. 
We may remark a rather good agreement for the Jahn-Teller energy $E_{JT}$ between 
our work and the optical data of Vallin {\em et al} for ZnS:Cr. \cite{Vallin70} However, our
estimate of the phonon frequency is much larger. A detailed analysis of this 
discrepancy is beyond the scope of the present work. We just remark that higher 
phonon frequencies would mean a more profound tendency towards the dynamic
Jahn-Teller effect. And there are indeed discussions (not for ZnS:Cr, however) whether 
the experimental data of GaN:Mn should be interpreted as a static \cite{Wolos04} or a 
dynamic \cite{Marcet06} Jahn-Teller effect. Such nonadiabatic effects    
can, however, not at all be treated in density functional based methods.

\section{Discussion and Conclusion}

Many ab-initio studies of Mn-doped GaN or Cr-doped ZnS, mostly based on the 
LSDA approximation, result in a half-metallic behavior. In those calculations, the Fermi 
level is located in a small 3$d$ band of majority spin within the middle of the gap of the 
host semi-conductor. \cite{Sanyal03,Kulatov02,McNorton08,Tablero06} We have shown,
that such a result is an artifact of the LSDA method and can be repaired by
taking into account the strong Coulomb correlation in the
$3d$ shell and the Jahn-Teller effect simultaneously. Allowing for a Jahn-Teller 
distortion only, but neglecting the Coulomb correlation leads to a tiny gap at the Fermi 
level.\cite{Luo05} And also the opposite procedure of taking into account the Coulomb 
correlation by the LSDA+$U$ method (i.e.\ the same method which we have used) 
but remaining in cubic symmetry is insufficient.\cite{Sandratskii04} The origin for this 
deficiency in the given case of a $d^4$ impurity is a threefold degenerate level of 
$t_{2g}$ symmetry at the Fermi level which is occupied with one electron only. It 
is sufficient to break the local cubic symmetry in the presence of a strong electron 
correlation to obtain insulating behavior with a gap of the order of 1 eV. Such an 
electronic structure is in agreement with the known experimental data for both compounds.
Using the LSDA+$U$ method we found the Jahn-Teller energy gain $E_{JT}$ in 
good agreement with known optical data.\cite{Wolos04,Kaminska79}

In addition to the ab-initio calculations, we developed a ligand field theory to reach 
a deeper understanding 
of the Jahn-Teller effect. It uses the $p$-$d$ hybridization between 3$d$ impurity and
ligands as principal origin of the crystal field splitting.\cite{Kuzian06} This hybridization 
is parametrized by the Harrison scheme\cite{Harrison80} and the lattice energy
as obtained from our ab-initio calculation is added. The resulting energy gain is very 
close to the ab-initio results but the ligand field theory allows in addition the determination 
of
the complete set of crystal field parameters. We find good agreement with the experimental
parameter set for GaN:Mn.\cite{Wolos04} The comparison is a little bit
speculative since the experimental data were obtained for Mn in wurtzite GaN. It turns out,
that the Mn impurity leads to an additional tetragonal Jahn-Teller distortion in addition to the
intrinsic trigonal deformation of the host lattice. And we find the parameters of this 
tetragonal distortion close to our results for zinc-blende GaN (which can be synthesized, but for 
which no measurements exist up to now). We have observed that our ligand field method
works better for GaN:Mn than for ZnS:Cr due to the very large cubic crystal field splitting 
$\Delta_q=1.4$ eV 
 in the former case (in contrast to  $\Delta_q=0.58$ eV for ZnS:Cr).
 
Our combined ab-initio and analytical study allows also a comparison
with the "classical" work on ZnS:Cr. The Jahn-Teller efect of that model compound
was already studied in the seventies in great detail by optical and EPR 
measurements. \cite{Vallin70,Vallin74, Kaminska79} 
We obtain an energy gain $E_{JT}$ very close to the experimental
data but higher phonon frequencies.  

Finally, we would like to discuss the importance of our results for spintronics applications.
First of all, the Jahn-Teller mechanism which we describe is not restricted to the two 
compounds of our study. It may occur in all cases where the impurity level is well
separated from the valence band but only partially filled. The Jahn-Teller effect leads to
insulating behavior which questions many previous ab-initio studies in the literature. In
the case of GaN:Mn all available information points to the stability of Mn$^{3+}$ which 
hinders an intrinsic hole doping by Mn. (Experimentally, one may find Mn$^{2+}$ in
electron doped samples, which is not well suited for spintronics applications either.) 
One still has the possibility to reach hole doping by a second impurity besides Mn 
(co-doping). That would allow the classical mechanism for ferromagnetism in 
diluted magnetic semiconductors with $S=2$ local moments at the Mn sites.

Our work was supported by a PICS project (No. 4767) and we thank Anatoli 
Stepanov and Andrey Titov for useful discussions.

\end{document}